\documentclass[aps,pra,twocolumn,showpacs,floatfix]{revtex4-1}  
\usepackage{graphicx}  
\usepackage{dcolumn}   
\usepackage{bm}        
\usepackage{amssymb}   
\usepackage{amsbsy}
\usepackage{amsmath}
\usepackage{color}
\usepackage{natbib}
\usepackage{hyperref}
\usepackage{float}

\usepackage[linesnumbered,ruled,lined,noline]{algorithm2e}

\hypersetup{
    colorlinks=true,
    linkcolor=blue,
    filecolor=blue,      
    urlcolor=blue,
    citecolor=blue,
    }

\newcommand{\sixj}[6]                      
{\left\{\begin{array}{ccc} #1&#2&#3 \\
#4&#5&#6 \end{array}\right\}}

\begin{document}
\title{
Portable GPU implementation of the WP-CCC ion-atom collisions code
}
\author{I.~B.~Abdurakhmanov$^{1}$}
\author{N.~W.~Antonio$^{2}$}
\author{M.~Cytowski$^{1}$}
\author{A.~S.~Kadyrov$^{2,3}$}
\affiliation{$^{1}$Pawsey Supercomputing Research Centre, 1 Bryce Avenue, Kensington, Western Australia 6151, Australia
}
\affiliation{$^{2}$Department of Physics and Astronomy, Curtin University, GPO Box U1987, Perth 6845, Australia
}
\affiliation{$^{3}$Institute of Nuclear Physics, Ulugbek, Tashkent 100214, Uzbekistan
}
\date{\today}

\begin{abstract}

In this manuscript we present our experience of porting the code used in the wave-packet convergent-close-coupling (WP-CCC) approach to run on NVIDIA V100 and AMD MI250X GPUs. The WP-CCC approach is a method used in the field of ion-atom collision physics to describe various processes such as elastic scattering, target excitation and electron-capture by the projectile. It has demonstrated its effectiveness in modelling collisions involving proton or bare ion projectiles with various atomic and molecular targets, especially those which can effectively be considered as one or two-electron systems. Such calculations find their application in computational atomic physics 
as well as in the modelling of fusion plasmas and in hadron therapy for cancer treatment.  The main computational cost of the method lies in the solution of an emerging set of coupled first-order differential equations. This involves implementing the standard Runge-Kutta method while varying the projectile position along multiple straight-line paths. At each projectile position several millions of matrix elements need to be calculated which is accomplished using the OpenACC programming model. Once these matrix elements are computed, the subsequent steps involve matrix inversion and multiplication with another matrix. To expedite these operations, a GPU-accelerated LAPACK routine, specialised for solving systems of linear equations, is employed. For AMD GPUs, this routine is accessible through the hipSOLVER library, while for NVIDIA GPUs, it can be obtained from the cuSOLVER library. The portability, performance and energy efficiency of the CPU-only code have been compared with the GPU-accelerated version running on AMD and NVIDIA GPUs.
The implementation of GPU-accelerated WP-CCC code  opens up avenues for exploring more sophisticated collision processes involving complex projectile and target structures, which were previously considered infeasible.


\end{abstract}   
\maketitle

\section{Introduction}

GPUs offer several advantages over CPUs for computational tasks due to their highly parallel architecture, featuring thousands of cores capable of executing tasks concurrently. This parallel processing power enables GPUs to handle large-scale computations with exceptional speed and efficiency, making them particularly well-suited for tasks such as graphics rendering, scientific simulations, and machine learning algorithms. Additionally, GPUs boast high memory bandwidth, enabling rapid data access and transfer, crucial for memory-intensive applications. Their energy efficiency, ability to offload tasks from CPUs, scalability with multiple GPUs, and cost-effectiveness further contribute to their appeal as powerful accelerators for a wide range of computational workloads.

Programming for GPUs used to be a complex task, requiring developers to write codes using low-level languages such as CUDA or OpenCL. These languages demanded a deep understanding of the underlying hardware architecture and intricate memory management techniques, making GPU programming inaccessible to many developers. Furthermore, optimizing code for GPUs often involved manual tuning and experimentation, which could be time-consuming and error-prone. As a result, the adoption of GPU acceleration in scientific computing was limited to a relatively small community of experts with specialized knowledge in parallel programming.

However, the invention of directive-based GPU offloading models such as OpenACC~\cite{OpenACC} and OpenMP~\cite{OpenMP} provided a significant boost for scientific computing by offering a simpler approach. 
It simplifies the process of GPU programming, without requiring extensive restructuring of the original CPU code.  Additionally, directive-based models provide portability across different GPU architectures and vendors, allowing code to run efficiently on a number of diverse hardware platforms without major modifications. 

While GPUs are highly effective in many computational tasks, they may demonstrate inefficiencies in certain scenarios. One notable limitation is their performance for tasks with irregular or sequential data dependencies, where parallelism is challenging to be exploited fully. In these cases, the overhead of coordinating threads and managing memory can outweigh the benefits of parallelism, leading to suboptimal performances compared to CPUs. Additionally, GPUs may struggle with tasks that involve frequent branching or conditional operations, as divergent thread execution can hinder parallel efficiency. Furthermore, GPU programming requires careful optimization and tuning to leverage the hardware effectively, and not all algorithms are easily parallelizable or well-suited for GPU acceleration. As a result, while GPUs offer significant advantages for many computational tasks, their efficiency may vary depending on the nature of the workload and the effectiveness of parallelization techniques employed.

In this work we present our experience and challenges encountered during the process
of GPU acceleration of the WP-CCC code. 
The CPU-based WP-CCC code, written in Fortran, employs MPI for parallelization across multiple nodes and OpenMP within each node~\cite{AKB16pra, ABKB18}.
The acronym WP-CCC stands for Wave-Packet Convergent Close-Coupling (WP-CCC) approach which is utilized to simulate ion-atom collisions.
The WP-CCC method offers a highly accurate and computationally efficient framework for simulating ion-atom collisions by coupling bound states and wave-packet pseudostates and iteratively refining the scattering solutions until convergence is achieved. This method allows researchers to study the dynamics of ion-atom interactions, including elastic scattering, excitation and electron capture processes, with high precision, making it a valuable tool for understanding fundamental atomic and molecular physics phenomena. 
Apart from the fundamental interest it is also useful in the modelling of fusion plasmas~\cite{Hemsworth_2009} and in hadron therapy for cancer treatment~\cite{Abril2013}.
It has demonstrated its effectiveness in studying collisions involving proton or bare ion projectiles with various atomic and molecular targets, especially those which can effectively be considered as one or two-electron systems~\cite{APAK24,PABK23,APABK22,AKFB13l}. However, the existing CPU-only code used for these computations experiences notable slowdowns as the complexity of the projectile and target increases. To address this limitation and extend the applicability of the method to more sophisticated collision processes, there is a pressing need to migrate the computations to GPUs. GPUs, consisting of thousands of cores,  can significantly enhance the computational efficiency and accelerate the calculations performed using the WP-CCC method.

The primary computational burden of the WP-CCC method lies in solving the resulting set of coupled first-order differential equations. This involves implementing the standard Runge-Kutta method while varying the projectile position along multiple straight-line paths.
In typical collision calculations, each path involves consideration of several thousand projectile positions, with the total number of paths typically ranging from 30 to 50.
At each projectile position, the computation involves evaluating several millions of matrix elements.
This task is handled using the OpenACC programming model, as the expression for matrix elements involves multiple summation operations, making it an excellent candidate for highly parallel execution.
Once these matrix elements are computed, the subsequent steps involve matrix inversion and multiplication with another matrix. To speed up these operations, a GPU-accelerated LAPACK routine, specialized for solving systems of linear equations, is employed. This library capitalizes on the parallel architecture of GPUs, enabling rapid and efficient computation of the necessary matrix transformations within the WP-CCC method. 
For AMD GPUs, this routine is accessible through the hipSOLVER library~\cite{hipsolver}, while for NVIDIA GPUs, it can be obtained from the cuSOLVER library~\cite{cusolver}. 

Thorough performance evaluation of the GPU-accelerated version of the code, executed on both AMD MI250X and NVIDIA V100 GPUs in comparison with the CPU-only code has been carried out. Through rigorous benchmarking and analysis, the computational efficiency gains achieved by migrating the calculations to GPUs have been quantified. The comparison covers various metrics such as execution time and scalability across different problem sizes. Notably, the GPU-accelerated implementation shows significant speedup, particularly pronounced when dealing with a case where the total scattering wave function is expanded in terms of a larger number of bound states and wave-packet pseudostates. 

The paper is set out as follows. Section II details 
the algorithms used in the WP-CCC code and their GPU implementation. 
The results of performance comparisons are presented in Section III. Finally in section IV we draw conclusions. 

\section{Brief description of the method and the GPU implementation}

The formalism for the WP-CCC approach has been outlined in Refs. ~\cite{AKB16pra, ABKB18}. Here, we only focus on the computational aspects of the method.
The WP-CCC method involves expanding the total scattering wave function of the collision system as a linear combination of basis functions. 
The basis functions consist of bound eigenstates and wave-packet pseudostates describing the projectile and the target atom.
By increasing the number of  these basis functions we are able to find convergence in the results obtained for
observables such as the cross sections and transition probabilities for each of the collision processes.. 

Skipping the mathematical derivations, we will transition directly to outlining the most time-consuming part of the code. 
It involves solving the following 
set of coupled first-order differential equations to determine the time-dependent coefficients, $\mathbf{F}$ and $\mathbf{G}$,
which represent direct scattering and electron transfer transitions, respectively:

\begin{eqnarray}
i v
\begin{bmatrix}
 \mathbf{I} & \mathbf{K} \\[10pt]
 \mathbf{K}^\dagger & \mathbf{I} 
\end{bmatrix}
\times
 \frac{d}{d z} 
\begin{bmatrix}
  \mathbf{F}  \\[10pt]
 \mathbf{G}
\end{bmatrix}
=
\begin{bmatrix}
  \mathbf{D}_\text{T} & \mathbf{Q}  \\[10pt]
  \mathbf{Q}^\dagger & \mathbf{D}_\text{P}
\end{bmatrix}
\times
\begin{bmatrix}
  \mathbf{F}  \\[10pt]
  \mathbf{G}
\end{bmatrix},
\end{eqnarray}
where $^\dagger$ denotes a complex conjugate operator and $\mathbf{I}$ is the identity matrix, 
$v$ is the speed of the projectile and $z$ is its position along the $z$-axis. 
Submatrices $\mathbf{K}$,  $\mathbf{D}_\text{T}$,  $\mathbf{D}_\text{P}$ and $\mathbf{Q}$ represent the overlap between 
the target and projectile states, 
direct target-target scattering, direct projectile-projectile scattering and electron-transfer, respectively.
Based on profiling of the CPU-based code, this part of the WP-CCC code 
proven to take more than 98\% of the total execution time in production runs.

In order to proceed with solving this system of equations we transform it to the following form:
\begin{eqnarray}
 \frac{d}{d z} 
 \mathbf{X} 
=
 \mathbf{M}\times\mathbf{X},
\label{DIFFEQ}
\end{eqnarray}
where a newly denoted solution vector $\mathbf{X}$ joins subvectors
$\mathbf{F}$ and  $\mathbf{G}$ 
\begin{eqnarray}
\mathbf{X}
=
\begin{bmatrix}
  \mathbf{F}  \\[10pt]
  \mathbf{G}
\end{bmatrix},
\end{eqnarray}
and 
\begin{eqnarray}
\mathbf{M}
=-\frac{i}{v}
\begin{bmatrix}
 \mathbf{I} & \mathbf{K} \\[10pt]
 \mathbf{K}^\dagger & \mathbf{I} 
\end{bmatrix}
^{-1}
\times
\begin{bmatrix}
  \mathbf{D}_\text{T} & \mathbf{Q}  \\[10pt]
  \mathbf{Q}^\dagger & \mathbf{D}_\text{P}
\end{bmatrix}
.
\label{INV}
\end{eqnarray}

A straightforward approach to calculating the matrix $\mathbf{M}$ based
on matrix inversion and subsequent scalar matrix-matrix
multiplication is prone to numerical errors. Instead, we
rely on the highly accurate ZGESV routine of LAPACK
library~\cite{lapack} and find the matrix M by solving the
system of linear equations. The GPU-accelerated version of 
ZGESV is also available.  For AMD GPUs, it can be accessed via the hipSOLVER library, 
whereas for NVIDIA GPUs, it is available through the cuSOLVER library.

It is worth noting that much of the computational complexities disappear when the projectile is negatively charged.
In this case, the projectile is unable to capture the target electron.
Consequently, the matrix $\mathbf{M}$ reduces to:
\begin{eqnarray}
\mathbf{M}
=
  \mathbf{D}_\text{T}
.
\label{NOINV}
\end{eqnarray}

The set of differential equations~\eqref{DIFFEQ} has parametrical dependence on the impact parameter, $\bm{b}$, which is
the perpendicular distance between the path of a projectile and the center of mass of the target. 
Therefore, it needs to be solved
while varying the projectile position along multiple straight-line paths characterised by the 
$z$-coordinate of the projectile and the impact parameter. 
The solution vector 
$\mathbf{X}$ is advanced along the predefined 
$z$-grid, by employing the standard fourth-order Runge-Kutta method~\cite{numerical_recipes}. 
When selecting the 
$z$-grid, we take into account the dynamics of the collision process. We aim for a denser distribution of points in the region where the projectile is closest to the target, ensuring greater accuracy in areas where the interaction between the particles is strongest. Conversely, we adopt a sparser distribution in the region where the projectile is furthest from the target, optimizing computational efficiency without compromising accuracy. To achieve this objective, the 
$z$-grid is constructed according to ${\rm zgrid}[i]=z_{\rm min} (z_{\rm max}/z_{\rm min})^{i/n_{\rm z}}$, 
${\rm zgrid}[0]=0$ and ${\rm zgrid}[-i]=-{\rm zgrid}[i]$, with $i$ ranging from 1 to $n_{\rm z}$. Here, the parameter 
$z_{\rm min}$ is utilized to adjust the density of the 
$z$-grid. In typical calculations, $z_{\rm min}$ is set to $10^{-3}$, $n_{\rm z}$ is set to 500, and $z_{\rm max}$
is set to 100 atomic units.

Algorithm~\ref{alg:myalgorithm} presents the pseudo code for the differential equation solver.  Lines 1 and 2 ensure that the initial boundary conditions are satisfied, 
meaning that the target atom is in the ground state before the collision occurs. 
At each step, the \textbf{for} loop implements the fourth-order Runge-Kutta method to advance the solution vector to the next 
$z$-position, $z_{\rm new}$.  This process involves computing four slopes, 
$\mathbf{K}_1$, $\mathbf{K}_2$, $\mathbf{K}_3$ and $\mathbf{K}_4$, which represent the derivative of the solution vector at different points within the step. 
That is achieved by evaluating the matrix $\mathbf{M}$ at two $z$-positions, $z_{\rm mid}$ and  $z_{\rm new}$.
During the overall cycle the matrix $\mathbf{M}$ gets evaluated $4n_{\rm z}+1$ times.
At the end of the $i$ loop the solution vector is evaluated at the position $z_{\rm max}$ which is used to calculate observables such as cross sections for various 
reaction processes.
In typical collision calculations, multiple paths characterized by various impact parameters 
$b$ need to be computed, with the total number typically ranging from 30 to 50. Each path involves the consideration of several thousand projectile positions and is handled by a separate MPI process. 
Consequently, communication between MPI processes is minimal, rendering the calculation embarrassingly parallel in terms of CPU cores.

\begin{algorithm}[t]
\caption{Algorithm for solving Eq. \eqref{DIFFEQ}}
\label{alg:myalgorithm}
\Indp\Indp\Indp\Indp\Indp\Indp\Indp\Indp\Indp\Indp\Indp\Indp
$\mathbf{X}[:]=0$\\
$\mathbf{X}[1]=1$\\
Computing $\mathbf{M}$ at ${\rm zgrid}[-n_{\rm z}]$\\
$\mathbf{M}_{\rm new}=\mathbf{M}({\rm zgrid}[-n_{\rm z}])$\\
\For{i=$-n_{\rm z}$ \KwTo $n_{\rm z}-1$}
{
\Indp\Indp\Indp\Indp\Indp\Indp\Indp\Indp\Indp\Indp\Indp\Indp\Indp\Indp\Indp\Indp\Indp\Indp\Indp\Indp\Indp\Indp\Indp\Indp 
$z={\rm zgrid}[i]$\\
$z_{\rm new}={\rm zgrid}[i+1]$\\
$z_{\rm mid}=(z+z_{\rm new})/2$\\
$h=z_{\rm new}-z$\\
$h_2=h/2$\\
$h_6=h/6$\\
$\mathbf{M}_z=\mathbf{M}_{\rm new}$\\
Computing $\mathbf{M}$ at $z_{\rm mid}$ and $z_{\rm new}$\\
$\mathbf{M}_{\rm mid}=\mathbf{M}(z_{\rm mid})$\\
$\mathbf{M}_{\rm new}=\mathbf{M}(z_{\rm new})$\\
$\mathbf{K}_1=\mathbf{M}_z\times\mathbf{X}$\\
$\mathbf{K}_2=\mathbf{M}_{\rm mid}\times\mathbf{X}+h_2 \mathbf{M}_{\rm mid}\times \mathbf{K}_1$\\
$\mathbf{K}_3=\mathbf{M}_{\rm mid}\times\mathbf{X}+h_2 \mathbf{M}_{\rm mid}\times \mathbf{K}_2$\\
$\mathbf{K}_4=\mathbf{M}_{\rm new}\times\mathbf{X}+h \mathbf{M}_{\rm new}\times \mathbf{K}_3$\\
$\mathbf{X}_{\rm new}=\mathbf{X}+h_6(\mathbf{K}_1+2\mathbf{K}_2+2\mathbf{K}_3+\mathbf{K}_4)$\\
$\mathbf{X}=\mathbf{X}_{\rm new}$\\
}
\end{algorithm}

The computation of the matrix $\mathbf{M}$ involves evaluating two structurally distinct types of matrix elements: direct matrix elements and overlap matrix elements. 
The electron-transfer matrix elements resemble overlap matrix elements 
and can be computed using a similar technique.
The final expression for the direct matrix elements, which constitute matrix  $\mathbf{D}_\text{T}$ is as follows:
\begin{align}
D^{}_{\alpha' \alpha} =&-\sum_{\lambda\mu}\frac{\sqrt{(2l_{\alpha}+1)(\lambda-\mu)}}{\sqrt{(2l_{\alpha'}+1)(\lambda+\mu)}}
C_{\lambda0l_{\alpha}0}^{l_{\alpha'}0}C_{\lambda\mu l_{\alpha}m_{\alpha}}^{l_{\alpha'}m_{\alpha'}}\nonumber \\ & \times
P_{\lambda}^{\mu}\left(\frac{z}{\sqrt{b^2+z^2}}\right)F_{n_{\alpha' }l_{\alpha' }n_{\alpha}l_{\alpha}\lambda}\left(\sqrt{b^2+z^2}\right),
\label{D1}
\end{align}
where $n_{\alpha}$, $l_{\alpha}$ and $m_{\alpha}$ are principal, orbital and magnetic quantum numbers of state $\alpha$,
$C_{\lambda\mu l_{\alpha}m_{\alpha}}^{l_{\alpha'}m_{\alpha'}}$ are Clebsch-Gordan coefficients,
$P_{\lambda}^{\mu}$ are Legendre polynomials and
 \begin{align}
F_{n_{\alpha' }l_{\alpha' }n_{\alpha}l_{\alpha}\lambda}\left(R\right)
=&\int_0^\infty dr r^2 f_{n_{\alpha' }l_{\alpha' }}(r)f_{n_{\alpha }l_{\alpha }}(r)\nonumber \\ & \times
\frac{\min(r,R)^{\lambda}}{\max(r,R)^{\lambda+1}}
,
\label{D2}
\end{align}
where $f_{n_{\alpha }l_{\alpha }}(r)$ is the radial function of the state $\alpha$.

The computation of another type of matrix elements, namely overlap matrix elements, relies on the following expression:
\begin{align}
K_{\beta\alpha}=&
\sum_{q'q}d_{m_{\beta}q'}^{l_{\beta}}(\theta)d_{m_{\alpha}q'}^{l_{\alpha}}(\theta)S_{n_{\beta}l_{\beta}q',n_{\beta}l_{\beta}q}(z).
\label{K1}
\end{align}
The Wigner rotation operator, $d_{m_{\alpha}q'}^{l_{\alpha}}(\theta)$, is employed to rotate the target frame to an angle of 
$\theta=\arccos(z/\sqrt{z^2+b^2})$  in order to transition to the collision frame.
The matrix elements in the target frame, $S_{n_{\beta}l_{\beta}q',n_{\beta}l_{\beta}q'}(z)$, are defined as:
\begin{align}
S_{n_{\beta}l_{\beta}q',n_{\beta}l_{\beta}q}(z)=&
\frac{i^{q-q'}(z^2+b^2)^{3/2}\sqrt{2l_{\alpha}+1}\sqrt{2l+1}}{16}\nonumber \\ & \times
\sqrt{\frac{(l_{\beta}-q')!(l_{\alpha}-q)!}{(l_{\beta}+q')!(l_{\alpha}+q)!}}\nonumber \\ & \times
\int_{1}^{\infty}d\eta
\int_{-1}^{1}d\tau(\eta^2-\tau^2)\exp\left(i  \frac{v z}{2} \eta \tau\right)\nonumber \\ & \times
f_{n_{\beta}l_{\beta}}\left(\frac{R(\eta+\tau)}{2}\right)
f_{n_{\alpha}l_{\alpha}}\left(\frac{R(\eta-\tau)}{2}\right)\nonumber \\ & \times
P_{l_{\beta}}^{q'}\left(\frac{\eta\tau+1}{\eta+\tau}\right)
P_{l_{\alpha}}^{q}\left(\frac{\eta\tau-1}{\eta-\tau}\right)\nonumber \\ & \times
J_{q-q'}\left(\frac{vb}{2}\sqrt{(\eta^2-1)(1-\tau^2)}\right),
\label{K2}
\end{align}
where $J_{q-q'}$ is the Bessel function of the first kind.

The computations of the $D^{}_{\alpha' \alpha} $ and $K_{\beta\alpha}$ matrix elements are offloaded to the GPU using the OpenACC programming model. To ensure optimal performance, we adhere to the OpenACC best practices by breaking down computations into smaller kernels rather than relying on a single large kernel. This approach enhances parallelism and load balancing, optimizes the memory access patterns, enables finer-grained parallelism, creates opportunities for targeted optimizations, and simplifies error isolation and debugging processes. Furthermore, based on our experience, this approach can also help reduce register pressure within the kernels.

Before computing the actual matrix elements, we first compute the necessary components in separate OpenACC kernels. These components include the quantities and functions
$\sqrt{2l+1}$,  $\sqrt{\lambda-\mu}$, $\sqrt{\lambda+\mu}$, $C_{\lambda\mu l_{\alpha}m_{\alpha}}^{l_{\alpha'}m_{\alpha'}}$, 
$P_{\lambda}^{\mu}\left(\frac{z}{\sqrt{b^2+z^2}}\right)$, $P_{l_{\beta}}^{q'}\left(\frac{\eta\tau+1}{\eta+\tau}\right)$, $P_{l_{\alpha}}^{q}\left(\frac{\eta\tau-1}{\eta-\tau}\right)$, 
$\exp\left(i  \frac{v z}{2} \eta \tau\right)$, 
$J_{q-q'}\left(\frac{vb}{2}\sqrt{(\eta^2-1)(1-\tau^2)}\right)$,
$f_{n_{\alpha }l_{\alpha }}(r)$ and $F_{n_{\alpha' }l_{\alpha' }n_{\alpha}l_{\alpha}\lambda}\left(R\right)$
for all possible index and argument values. To enhance parallelism, multi-dimensional indices, such as the six-dimensional  $n_{\alpha' }l_{\alpha' }n_{\alpha}l_{\alpha}\lambda$
in $F_{n_{\alpha' }l_{\alpha' }n_{\alpha}l_{\alpha}\lambda}\left(R\right)$, are converted into one dimension. 
This process is equivalent to implementing the \textbf{collapse} directive, which combines a specified number of indices into a single dimension. However, we have observed that our manual approach yields faster code execution and enables the storage of arrays in a more compact form with better memory access.
Whenever feasible, we implement gang 
$+$ vector levels of parallelism for these unified indices.
Moreover, the following components are precomputed and stored in the GPU global memory prior to proceeding with the equation solver: $\sqrt{2l+1}$,  $\sqrt{\lambda-\mu}$, $\sqrt{\lambda+\mu}$, $C_{\lambda\mu l_{\alpha}m_{\alpha}}^{l_{\alpha'}m_{\alpha'}}$,
$f_{n_{\alpha }l_{\alpha }}(r)$ and $F_{n_{\alpha' }l_{\alpha' }n_{\alpha}l_{\alpha}\lambda}\left(R\right)$, as they are
independent of either $z$ or $b$ characterising the projectile path.
The arguments $R$ and $r$ are discretized into a predefined radial grid, $r$-grid, spanning the range from 0 to 100 atomic units, typically comprising up to 10,000 points. The grid is denser around the origin to capture finer details in that region. The same grid is also used in the computation of the integral
in Eq.~\eqref{D2}. 
The integrals over $\eta$ and $\tau$ in Eq.~\eqref{K2} are computed using the Gauss-Laguerre and Gauss-Legendre quadrature points.
When computing the radial wave function $f_{n_{\alpha }l_{\alpha }}$ or the integral $F_{n_{\alpha' }l_{\alpha' }n_{\alpha}l_{\alpha}\lambda}$
at arguments beyond the $r$-grid point, such as at $\sqrt{b^2+z^2}$ as shown in Eq.~\eqref{D1}, we employ a five-point polynomial interpolation method.
With all necessary components stored in a global GPU memory the last step is to compute matrix  $\mathbf{D}_\text{T}$ and 
$\mathbf{K}$,  where the pair
of indices $\alpha' \alpha$ and $\alpha\beta$ are converted into the unifying index which ranges from one to several millions in production calculations. 
This index is utilized within the gang $+$ vector levels of OpenACC parallelism.
The computations of other submatrices, namely $\mathbf{D}_\text{P}$ and 
$\mathbf{Q}$, are carried out using a similar approach.

\section{Performance analysis}

For our performance analysis, we utilise the prototype system of proton collisions with atomic hydrogen. 
This system represents the simplest collision scenario, 
comprising of only three particles: two protons and an electron. Despite its simplicity, 
it encompasses a wide range of reaction channels that are typical of ion-atom collisions.
We consider two types of calculations, the so called one-centre (1C) and two-centre (2C) calculations.
The 1C calculations are simpler in nature, requiring computations of only one type of matrix elements: direct matrix elements. 
They are particularly relevant for collisions where the projectile is negatively charged, where electron capture is impossible. While also applicable to collisions involving positively charged projectiles, they are limited to determining total electron-loss cross sections and lack the capability to distinguish between the electron-capture and ionization processes.
The 2C calculations utilize the entire codebase and possess significantly broader applications.

We select an impact energy of 100 keV for the projectile. That is the energy where all reaction channels are approximately equally significant.
The $z$-grid describing the projectile path spans from $-300$ to $300$ atomic units, containing 6001 points. To accommodate the number of all available GPUs within the Setonix GPU node, each calculation simultaneously considers eight values of the impact parameter, or in other words, eight paths.

\begin{figure}[ht]
\includegraphics[width=\columnwidth]{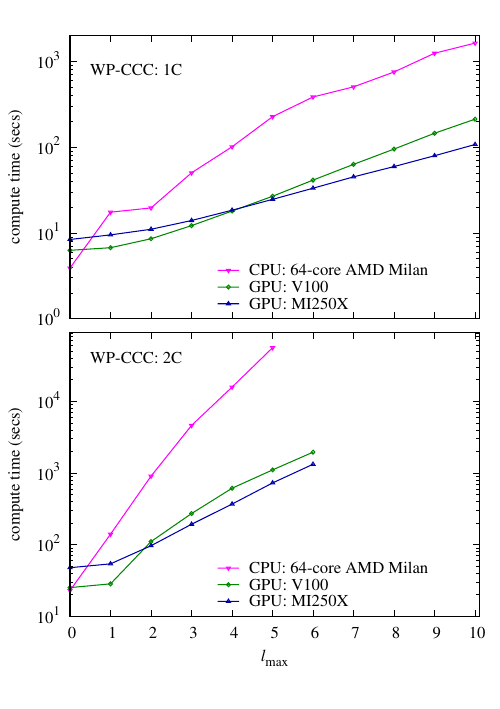}
\caption{The compute time of the WP-CCC code in both 1C and 2C modes is depicted as a function of the maximum allowed angular momentum, 
$l_{\rm max}$, a parameter that governs the size of the set of differential equations (Eq. ~\eqref{DIFFEQ}). The performance of the CPU-only code has been compared with that of the GPU-accelerated version, executed respectively on the V100 NVIDIA GPU and the MI250X AMD GPU.
}
\label{FIG1}
\end{figure}

\begin{figure}[ht]
\includegraphics[width=\columnwidth]{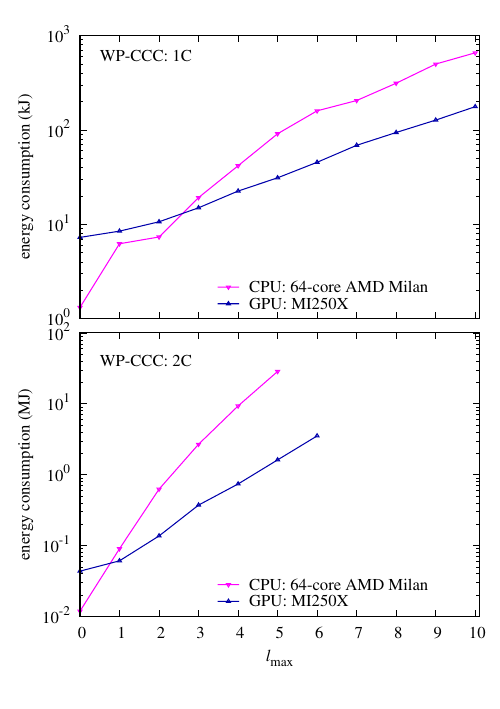}
\caption{The energy consumption of the WP-CCC code in both 1C and 2C modes is depicted as a function of the maximum allowed angular momentum, 
$l_{\rm max}$, a parameter that governs the size of the set of differential equations (Eq. ~\eqref{DIFFEQ}). The energy efficiency of the CPU-only code has been compared with that of the GPU-accelerated version, executed respectively on the V100 NVIDIA GPU and the MI250X AMD GPU.
}
\label{FIG2}
\end{figure}

\begin{figure}[ht]
\includegraphics[width=\columnwidth]{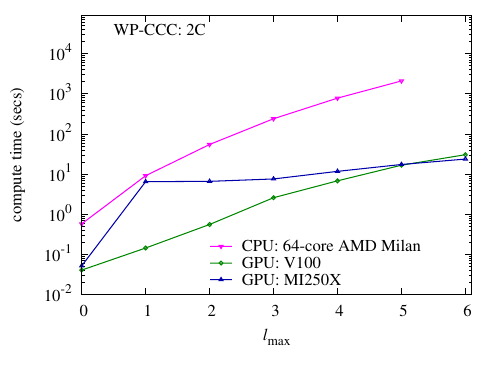}
\caption{(color online)
The compute time of linear equation solver within WP-CCC code in 2C mode is depicted as a function of the maximum allowed angular momentum, 
$l_{\rm max}$, a parameter that governs the size of the set of differential equations (Eq. ~\eqref{DIFFEQ}). The performance of the CPU-version of LAPACK ZGESV has been compared with the GPU-accelerated versions of ZGESV from cuSOLVER and hipSOLVER libraries.
}
\label{FIG3}
\end{figure}

We execute our GPU-accelerated code on two different systems: the Setonix GPU node, which comprises 64 AMD EPYC 7A53 CPU "Trento" cores and 8 AMD MI250X GCDs~\cite{SETONIX}, and the Gadi NCI GPU node, which consists of 48 Intel Xeon Scalable 'Cascade Lake' CPUs and 4 NVIDIA V100 GPUs. However, during the GPU job execution, only the number of CPU cores equal to the number of GPUs is utilized. The remaining CPU cores remain idle.
The GPU-accelerated code running on Setonix MI250X GPUs is compiled using the Cray Fortran compiler provided with the CPE 23.03, while on NVIDIA V100 GPUs, it is compiled using NVFortran from the PGI 19.7 suite.

For CPU-only jobs based on the MPI $+$ OpenMP parallelism model, we utilize the Setonix CPU node, which has 128 AMD EPYC 7763 CPU "Milan" cores.
Each of these nodes contain 2 CPU sockets, enabling a maximum of 64 OpenMP cores. Consequently, the CPU-based WP-CCC code can utilize up to 64 CPU cores concurrently to compute results for a single impact parameter.

Figure~\ref{FIG1} illustrates the performance and scalability of the WP-CCC CPU and GPU codes as the size of the problem increases by incrementing the maximum allowed angular momentum quantum numbers from 0 to 10.
In the 1C case, the size of the underlying set of first-order differential equations gradually grows from 42 at $l_{\max}=0$ to 4257 at $l_{\max}=10$.
In the 2C case, it increases from 60 at $l_{\max}=0$ to 1910 at $l_{\max}=5$.
We compare the total compute time of the WP-CCC code required to obtain results for one impact parameter when executed on 64 CPU "Milan" cores, 1 NVIDIA V100 GPU, and 1 AMD MI250X GCD.
In both 1C and 2C cases, we do not observe significant acceleration of the code when the problem size is small. However, as the problem size increases, the GPU versions of the code demonstrate increasingly faster performance compared to the CPU code.
Specifically, in the 1C case, when $l_{\max}=10$, the GPU code executed on the V100 GPU is 7.6 times faster, 
while on the MI250X GPU, it is 15.17 times faster compared to the computational performance 
achieved using the 64 "Milan" cores of a single CPU socket within the 128-core Setonix CPU node.
In the 2C case, when $l_{\max}=5$,  the speedup factors are even larger, reaching 77 for the AMD GPU and 51 for the V100 GPU.
At $l_{\max}=6$, our 2C CPU job exceeds the 24-hour walltime limit and does not finish.

The significant speedups achieved by the GPU code also translate into reduced energy consumption. 
This comparison was achieved by using HPE Cray EX Application Task Orchestration and Management (ATOM) energy reports 
which are integrated with SLURM workload manager on Setonix.
In Figure~\ref{FIG2}, we examine the energy consumption of the calculations depicted in Figure~\ref{FIG1}.
For the largest calculations shown in the 1C case ($l_{\max}=10$), the MI250X GPU consumed 3.7 times less energy than the 64-core CPU node.
In the 2C case ($l_{\max}=5$), the MI250X GPU is even more efficient, consuming 17 times less energy.
Furthermore, it is important to note that a single 128-core CPU node can only process 2 impact parameters at a time, while the AMD MI250X GPU can process 8 impact parameters simultaneously. This factor alone makes the 1C GPU code 15 times more energy efficient and the 2C GPU code 68 times more energy efficient.

Lastly, Figure~\ref{FIG3} presents a comparison of the total time spent on the linear equation solver within the WP-CCC code in the 2C case. It is evident that the GPU-accelerated linear equation solvers from the hipSOLVER and cuSOLVER libraries outperform the CPU-based multithreaded equation solver from the LAPACK library by two orders of magnitude. When comparing the performance of cuSOLVER and hipSOLVER, it becomes apparent that cusolver's performance is more stable for smaller problem sizes.

\section{Conclusion}

In conclusion, this manuscript has detailed our experience in porting the WP-CCC code to run efficiently on both NVIDIA V100 and AMD MI250X GPUs. 

Our comparisons have shown significant performance improvements and energy efficiencies with GPU acceleration.  For the largest performed calculations,  in the 1C case, the GPU code executed on the V100 GPU was 7.6 times faster, while on the MI250X GPU, it was 15.17 times faster than on the 64-core CPU node. In the 2C case, speedup factors were even larger, reaching 77 for the AMD GPU and 51 for the V100 GPU. 

Moreover, our energy consumption evaluations revealed significant gains. For identical computational tasks, the MI250X GPU consumed 15 times less energy in the 1C case and a remarkable 68 times less energy in the 2C case compared to CPU-only executions. Overall, our findings demonstrate substantial benefits of GPU acceleration in improving both performance and energy efficiency in WP-CCC calculations.

Furthermore, the implementation of GPU-accelerated WP-CCC code paves the way for exploring more intricate collision processes involving complex projectile and target structures that were previously impossible. This progress holds the potential to further advance research in ion-atom collision physics and related disciplines.

\begin{acknowledgments}
This work was supported by the Australian Research Council, the National Computing Infrastructure
and the resources provided by the Pawsey Supercomputing Research Centre with funding from the Australian Government 
and the Government of Western Australia and the Pawsey Centre for Extreme Scale Readiness (PaCER). 
N.W.A. acknowledges support through an Australian Government Research Training Program Scholarship.

\end{acknowledgments} 


\begin{thebibliography}{15}%
\makeatletter
\providecommand \@ifxundefined [1]{%
 \@ifx{#1\undefined}
}%
\providecommand \@ifnum [1]{%
 \ifnum #1\expandafter \@firstoftwo
 \else \expandafter \@secondoftwo
 \fi
}%
\providecommand \@ifx [1]{%
 \ifx #1\expandafter \@firstoftwo
 \else \expandafter \@secondoftwo
 \fi
}%
\providecommand \natexlab [1]{#1}%
\providecommand \enquote  [1]{``#1''}%
\providecommand \bibnamefont  [1]{#1}%
\providecommand \bibfnamefont [1]{#1}%
\providecommand \citenamefont [1]{#1}%
\providecommand \href@noop [0]{\@secondoftwo}%
\providecommand \href [0]{\begingroup \@sanitize@url \@href}%
\providecommand \@href[1]{\@@startlink{#1}\@@href}%
\providecommand \@@href[1]{\endgroup#1\@@endlink}%
\providecommand \@sanitize@url [0]{\catcode `\\12\catcode `\$12\catcode
  `\&12\catcode `\#12\catcode `\^12\catcode `\_12\catcode `\%12\relax}%
\providecommand \@@startlink[1]{}%
\providecommand \@@endlink[0]{}%
\providecommand \url  [0]{\begingroup\@sanitize@url \@url }%
\providecommand \@url [1]{\endgroup\@href {#1}{\urlprefix }}%
\providecommand \urlprefix  [0]{URL }%
\providecommand \Eprint [0]{\href }%
\providecommand \doibase [0]{http://dx.doi.org/}%
\providecommand \selectlanguage [0]{\@gobble}%
\providecommand \bibinfo  [0]{\@secondoftwo}%
\providecommand \bibfield  [0]{\@secondoftwo}%
\providecommand \translation [1]{[#1]}%
\providecommand \BibitemOpen [0]{}%
\providecommand \bibitemStop [0]{}%
\providecommand \bibitemNoStop [0]{.\EOS\space}%
\providecommand \EOS [0]{\spacefactor3000\relax}%
\providecommand \BibitemShut  [1]{\csname bibitem#1\endcsname}%
\let\auto@bib@innerbib\@empty
\bibitem [{\citenamefont {Chandrasekaran}\ and\ \citenamefont
  {Juckeland}(2017)}]{OpenACC}%
  \BibitemOpen
  \bibfield  {author} {\bibinfo {author} {\bibfnamefont {S.}~\bibnamefont
  {Chandrasekaran}}\ and\ \bibinfo {author} {\bibfnamefont {G.}~\bibnamefont
  {Juckeland}},\ }\href@noop {} {\emph {\bibinfo {title} {OpenACC for
  Programmers: Concepts and Strategies}}}\ (\bibinfo  {publisher}
  {Addison-Wesley Professional},\ \bibinfo {address} {Boston, MA},\ \bibinfo
  {year} {2017})\BibitemShut {NoStop}%
\bibitem [{\citenamefont {Deakin}\ and\ \citenamefont
  {Mattson}(2023)}]{OpenMP}%
  \BibitemOpen
  \bibfield  {author} {\bibinfo {author} {\bibfnamefont {T.}~\bibnamefont
  {Deakin}}\ and\ \bibinfo {author} {\bibfnamefont {T.~G.}\ \bibnamefont
  {Mattson}},\ }\href@noop {} {\emph {\bibinfo {title} {Programming Your GPU
  with OpenMP: Performance Portability for GPUs}}}\ (\bibinfo  {publisher} {MIT
  Press},\ \bibinfo {year} {2023})\BibitemShut {NoStop}%
\bibitem [{\citenamefont {Abdurakhmanov}\ \emph {et~al.}(2016)\citenamefont
  {Abdurakhmanov}, \citenamefont {Kadyrov},\ and\ \citenamefont
  {Bray}}]{AKB16pra}%
  \BibitemOpen
  \bibfield  {author} {\bibinfo {author} {\bibfnamefont {I.~B.}\ \bibnamefont
  {Abdurakhmanov}}, \bibinfo {author} {\bibfnamefont {A.~S.}\ \bibnamefont
  {Kadyrov}}, \ and\ \bibinfo {author} {\bibfnamefont {I.}~\bibnamefont
  {Bray}},\ }\href {\doibase 10.1103/PhysRevA.94.022703} {\bibfield  {journal}
  {\bibinfo  {journal} {Phys. Rev. A}\ }\textbf {\bibinfo {volume} {94}},\
  \bibinfo {pages} {022703} (\bibinfo {year} {2016})}\BibitemShut {NoStop}%
\bibitem [{\citenamefont {Abdurakhmanov}\ \emph {et~al.}(2018)\citenamefont
  {Abdurakhmanov}, \citenamefont {Bailey}, \citenamefont {Kadyrov},\ and\
  \citenamefont {Bray}}]{ABKB18}%
  \BibitemOpen
  \bibfield  {author} {\bibinfo {author} {\bibfnamefont {I.~B.}\ \bibnamefont
  {Abdurakhmanov}}, \bibinfo {author} {\bibfnamefont {J.~J.}\ \bibnamefont
  {Bailey}}, \bibinfo {author} {\bibfnamefont {A.~S.}\ \bibnamefont {Kadyrov}},
  \ and\ \bibinfo {author} {\bibfnamefont {I.}~\bibnamefont {Bray}},\ }\href
  {\doibase 10.1103/PhysRevA.97.032707} {\bibfield  {journal} {\bibinfo
  {journal} {Phys. Rev. A}\ }\textbf {\bibinfo {volume} {97}},\ \bibinfo
  {pages} {032707} (\bibinfo {year} {2018})}\BibitemShut {NoStop}%
\bibitem [{\citenamefont {Hemsworth}\ \emph {et~al.}(2009)\citenamefont
  {Hemsworth}, \citenamefont {Decamps}, \citenamefont {Graceffa}, \citenamefont
  {Schunke}, \citenamefont {Tanaka}, \citenamefont {Dremel}, \citenamefont
  {Tanga}, \citenamefont {Esch}, \citenamefont {Geli}, \citenamefont {Milnes},
  \citenamefont {Inoue}, \citenamefont {Marcuzzi}, \citenamefont {Sonato},\
  and\ \citenamefont {Zaccaria}}]{Hemsworth_2009}%
  \BibitemOpen
  \bibfield  {author} {\bibinfo {author} {\bibfnamefont {R.}~\bibnamefont
  {Hemsworth}}, \bibinfo {author} {\bibfnamefont {H.}~\bibnamefont {Decamps}},
  \bibinfo {author} {\bibfnamefont {J.}~\bibnamefont {Graceffa}}, \bibinfo
  {author} {\bibfnamefont {B.}~\bibnamefont {Schunke}}, \bibinfo {author}
  {\bibfnamefont {M.}~\bibnamefont {Tanaka}}, \bibinfo {author} {\bibfnamefont
  {M.}~\bibnamefont {Dremel}}, \bibinfo {author} {\bibfnamefont
  {A.}~\bibnamefont {Tanga}}, \bibinfo {author} {\bibfnamefont {H.~D.}\
  \bibnamefont {Esch}}, \bibinfo {author} {\bibfnamefont {F.}~\bibnamefont
  {Geli}}, \bibinfo {author} {\bibfnamefont {J.}~\bibnamefont {Milnes}},
  \bibinfo {author} {\bibfnamefont {T.}~\bibnamefont {Inoue}}, \bibinfo
  {author} {\bibfnamefont {D.}~\bibnamefont {Marcuzzi}}, \bibinfo {author}
  {\bibfnamefont {P.}~\bibnamefont {Sonato}}, \ and\ \bibinfo {author}
  {\bibfnamefont {P.}~\bibnamefont {Zaccaria}},\ }\href {\doibase
  10.1088/0029-5515/49/4/045006} {\bibfield  {journal} {\bibinfo  {journal}
  {Nuclear Fusion}\ }\textbf {\bibinfo {volume} {49}},\ \bibinfo {pages}
  {045006} (\bibinfo {year} {2009})}\BibitemShut {NoStop}%
\bibitem [{\citenamefont {Abril}\ \emph {et~al.}(2013)\citenamefont {Abril},
  \citenamefont {Garcia-Molina}, \citenamefont {de~Vera}, \citenamefont
  {Kyriakou},\ and\ \citenamefont {Emfietzoglou}}]{Abril2013}%
  \BibitemOpen
  \bibfield  {author} {\bibinfo {author} {\bibfnamefont {I.}~\bibnamefont
  {Abril}}, \bibinfo {author} {\bibfnamefont {R.}~\bibnamefont
  {Garcia-Molina}}, \bibinfo {author} {\bibfnamefont {P.}~\bibnamefont
  {de~Vera}}, \bibinfo {author} {\bibfnamefont {I.}~\bibnamefont {Kyriakou}}, \
  and\ \bibinfo {author} {\bibfnamefont {D.}~\bibnamefont {Emfietzoglou}},\
  }\href {\doibase 10.1016/B978-0-12-396455-7.00006-6} {\bibfield  {journal}
  {\bibinfo  {journal} {Adv. Quantum Chem.}\ }\textbf {\bibinfo {volume}
  {65}},\ \bibinfo {pages} {129} (\bibinfo {year} {2013})}\BibitemShut
  {NoStop}%
\bibitem [{\citenamefont {Antonio}\ \emph {et~al.}(2024)\citenamefont
  {Antonio}, \citenamefont {Plowman}, \citenamefont {Abdurakhmanov},\ and\
  \citenamefont {Kadyrov}}]{APAK24}%
  \BibitemOpen
  \bibfield  {author} {\bibinfo {author} {\bibfnamefont {N.~W.}\ \bibnamefont
  {Antonio}}, \bibinfo {author} {\bibfnamefont {C.~T.}\ \bibnamefont
  {Plowman}}, \bibinfo {author} {\bibfnamefont {I.~B.}\ \bibnamefont
  {Abdurakhmanov}}, \ and\ \bibinfo {author} {\bibfnamefont {A.~S.}\
  \bibnamefont {Kadyrov}},\ }\href {\doibase 10.1103/PhysRevA.109.012817}
  {\bibfield  {journal} {\bibinfo  {journal} {Phys. Rev. A}\ }\textbf {\bibinfo
  {volume} {109}},\ \bibinfo {pages} {012817} (\bibinfo {year}
  {2024})}\BibitemShut {NoStop}%
\bibitem [{\citenamefont {Plowman}\ \emph {et~al.}(2023)\citenamefont
  {Plowman}, \citenamefont {Abdurakhmanov}, \citenamefont {Bray},\ and\
  \citenamefont {Kadyrov}}]{PABK23}%
  \BibitemOpen
  \bibfield  {author} {\bibinfo {author} {\bibfnamefont {C.~T.}\ \bibnamefont
  {Plowman}}, \bibinfo {author} {\bibfnamefont {I.~B.}\ \bibnamefont
  {Abdurakhmanov}}, \bibinfo {author} {\bibfnamefont {I.}~\bibnamefont {Bray}},
  \ and\ \bibinfo {author} {\bibfnamefont {A.~S.}\ \bibnamefont {Kadyrov}},\
  }\href {\doibase 10.1103/PhysRevA.107.032824} {\bibfield  {journal} {\bibinfo
   {journal} {Phys. Rev. A}\ }\textbf {\bibinfo {volume} {107}},\ \bibinfo
  {pages} {032824} (\bibinfo {year} {2023})}\BibitemShut {NoStop}%
\bibitem [{\citenamefont {Alladustov}\ \emph {et~al.}(2022)\citenamefont
  {Alladustov}, \citenamefont {Plowman}, \citenamefont {Abdurakhmanov},
  \citenamefont {Bray},\ and\ \citenamefont {Kadyrov}}]{APABK22}%
  \BibitemOpen
  \bibfield  {author} {\bibinfo {author} {\bibfnamefont {S.~U.}\ \bibnamefont
  {Alladustov}}, \bibinfo {author} {\bibfnamefont {C.~T.}\ \bibnamefont
  {Plowman}}, \bibinfo {author} {\bibfnamefont {I.~B.}\ \bibnamefont
  {Abdurakhmanov}}, \bibinfo {author} {\bibfnamefont {I.}~\bibnamefont {Bray}},
  \ and\ \bibinfo {author} {\bibfnamefont {A.~S.}\ \bibnamefont {Kadyrov}},\
  }\href {\doibase 10.1103/PhysRevA.106.062819} {\bibfield  {journal} {\bibinfo
   {journal} {Phys. Rev. A}\ }\textbf {\bibinfo {volume} {106}},\ \bibinfo
  {pages} {062819} (\bibinfo {year} {2022})}\BibitemShut {NoStop}%
\bibitem [{\citenamefont {Abdurakhmanov}\ \emph {et~al.}(2013)\citenamefont
  {Abdurakhmanov}, \citenamefont {Kadyrov}, \citenamefont {Fursa},\ and\
  \citenamefont {Bray}}]{AKFB13l}%
  \BibitemOpen
  \bibfield  {author} {\bibinfo {author} {\bibfnamefont {I.~B.}\ \bibnamefont
  {Abdurakhmanov}}, \bibinfo {author} {\bibfnamefont {A.~S.}\ \bibnamefont
  {Kadyrov}}, \bibinfo {author} {\bibfnamefont {D.~V.}\ \bibnamefont {Fursa}},
  \ and\ \bibinfo {author} {\bibfnamefont {I.}~\bibnamefont {Bray}},\ }\href
  {\doibase 10.1103/PhysRevLett.111.173201} {\bibfield  {journal} {\bibinfo
  {journal} {Phys. Rev. Lett.}\ }\textbf {\bibinfo {volume} {111}},\ \bibinfo
  {pages} {173201} (\bibinfo {year} {2013})}\BibitemShut {NoStop}%
\bibitem [{hip()}]{hipsolver}%
  \BibitemOpen
  \href@noop {} {\enquote {\bibinfo {title} {hisolver: A gpu-accelerated
  library for dense and batched linear algebra subroutines},}\ }\bibinfo
  {howpublished}
  {https://github.com/ROCmSoftwarePlatform/hipsolver}\BibitemShut {NoStop}%
\bibitem [{cus()}]{cusolver}%
  \BibitemOpen
  \href@noop {} {\enquote {\bibinfo {title} {cusolver: A gpu-accelerated
  library for dense and batched linear algebra subroutines},}\ }\bibinfo
  {howpublished} {https://developer.nvidia.com/cusolver}\BibitemShut {NoStop}%
\bibitem [{\citenamefont {Anderson}\ \emph {et~al.}(1992)\citenamefont
  {Anderson}, \citenamefont {Bai}, \citenamefont {Bischof}, \citenamefont
  {Demmel}, \citenamefont {Dongarra}, \citenamefont {Croz}, \citenamefont
  {Greenbaum}, \citenamefont {Hammarling}, \citenamefont {McKenney},
  \citenamefont {Ostrouchov}, ,\ and\ \citenamefont {Sorensen}}]{lapack}%
  \BibitemOpen
  \bibfield  {author} {\bibinfo {author} {\bibfnamefont {E.}~\bibnamefont
  {Anderson}}, \bibinfo {author} {\bibfnamefont {Z.}~\bibnamefont {Bai}},
  \bibinfo {author} {\bibfnamefont {C.}~\bibnamefont {Bischof}}, \bibinfo
  {author} {\bibfnamefont {J.}~\bibnamefont {Demmel}}, \bibinfo {author}
  {\bibfnamefont {J.}~\bibnamefont {Dongarra}}, \bibinfo {author}
  {\bibfnamefont {J.~D.}\ \bibnamefont {Croz}}, \bibinfo {author}
  {\bibfnamefont {A.}~\bibnamefont {Greenbaum}}, \bibinfo {author}
  {\bibfnamefont {S.}~\bibnamefont {Hammarling}}, \bibinfo {author}
  {\bibfnamefont {A.}~\bibnamefont {McKenney}}, \bibinfo {author}
  {\bibfnamefont {S.}~\bibnamefont {Ostrouchov}}, , \ and\ \bibinfo {author}
  {\bibfnamefont {D.}~\bibnamefont {Sorensen}},\ }\href@noop {} {\emph
  {\bibinfo {title} {{LAPACK Users's guide}}}}\ (\bibinfo  {publisher} {Society
  for Industrial and Applied Mathematics},\ \bibinfo {address} {Philadelphia,
  USA},\ \bibinfo {year} {1992})\BibitemShut {NoStop}%
\bibitem [{\citenamefont {Press}\ \emph {et~al.}(2007)\citenamefont {Press},
  \citenamefont {Teukolsky}, \citenamefont {Vetterling},\ and\ \citenamefont
  {Flannery}}]{numerical_recipes}%
  \BibitemOpen
  \bibfield  {author} {\bibinfo {author} {\bibfnamefont {W.~H.}\ \bibnamefont
  {Press}}, \bibinfo {author} {\bibfnamefont {S.~A.}\ \bibnamefont
  {Teukolsky}}, \bibinfo {author} {\bibfnamefont {W.~T.}\ \bibnamefont
  {Vetterling}}, \ and\ \bibinfo {author} {\bibfnamefont {B.~P.}\ \bibnamefont
  {Flannery}},\ }\href@noop {} {\emph {\bibinfo {title} {Numerical Recipes: The
  Art of Scientific Computing}}},\ \bibinfo {edition} {3rd}\ ed.\ (\bibinfo
  {publisher} {Cambridge University Press},\ \bibinfo {address} {Cambridge,
  UK},\ \bibinfo {year} {2007})\BibitemShut {NoStop}%
\bibitem [{SET()}]{SETONIX}%
  \BibitemOpen
  \href {\doibase 10.48569/18sb-8s43} {\enquote {\bibinfo {title} {Setonix
  supercomputing research centre},}\ }\bibinfo {howpublished}
  {\url{https://doi.org/10.48569/18sb-8s43}}\BibitemShut {NoStop}%
\end{thebibliography}
%

\end{document}